\documentclass[aps,pre,preprint]{revtex4-1}
\usepackage{graphicx}
\draft
\usepackage{dcolumn}
\usepackage{bm}

\begin{document}

\title{Intrinsic Noise Induced Coherence Resonance in a Glow discharge Plasma}
 \author{Pankaj Kumar Shaw}
\email{pankaj.shaw@saha.ac.in} \affiliation{Plasma Physics
  Division, Saha Institute of Nuclear Physics, 1/AF, Bidhannagar,
  Kolkata -700064, India.}
   \author{Debajyoti Saha}
\email{debajyoti.saha@saha.ac.in} \affiliation{Plasma Physics
  Division, Saha Institute of Nuclear Physics, 1/AF, Bidhannagar,
  Kolkata -700064, India.}
   \author{Sabuj Ghosh}
\email{sabuj.ghosh@saha.ac.in} \affiliation{Plasma Physics
  Division, Saha Institute of Nuclear Physics, 1/AF, Bidhannagar,
  Kolkata -700064, India.}
   \author{M. S. Janaki}
\email{ms.janaki@saha.ac.in} \affiliation{Plasma Physics
  Division, Saha Institute of Nuclear Physics, 1/AF, Bidhannagar,
  Kolkata -700064, India.}
  \author{A. N. Sekar Iyengar}
\email{ansekar.iyengar@saha.ac.in} \affiliation{Plasma Physics
  Division, Saha Institute of Nuclear Physics, 1/AF, Bidhannagar,
  Kolkata -700064, India.}

\begin{abstract}
  Experimental evidence of intrinsic noise induced coherence resonance in a glow discharge plasma is being reported. Initially the system is started  at a discharge voltage (DV) where it exhibited fixed point dynamics, and then with the subsequent increase in the DV  spikes were excited which were few in number and with further increase of DV the number of spikes as well as their regularity increased. The regularity in the interspike interval of the  spikes is estimated using normalized variance (NV).  Coherence resonance was determined using normalized variance curve and also corroborated by Hurst exponent and power spectrum plots.  We show that the regularity of the excitable spikes in the floating potential fluctuation increases with the increase in the DV, upto  a particular value of DV.  Using a Wiener filter, we separated the noise component which was observed to increase with DV and hence conjectured that noise can be playing an important role in the generation of the coherence resonance.   From an anharmonic oscillator equation describing ion acoustic oscillations, we have been able to obtain a FHN like model which has been used to understand the excitable dynamics of glow discharge plasma in the presence of noise. The numerical results agree quite well with the experimental results.
\end{abstract}

\maketitle
\section{Introduction}
\label{section:introduction}
Noise is omnipresent in all natural systems and plays a beneficial role in the dynamics of  nonlinear systems yielding interesting results~\cite{lopes,lindner}. There are many studies which have revealed that noise can  play a constructive role like noise induced order in chaotic dynamics~\cite{matsu}, stochastic ~\cite{benzi,tang,zhou,gluc}  and coherence resonances~\cite{simon,pikovsky,kiss,santos}. Noise can be divided into two categories: extrinsic and intrinsic. The extrinsic noise can  originate  from the environment~\cite{silva} and  an external noise generator in experimental systems,  whereas intrinsic noise is generated due to an interplay between the components of the systems and could be of small amplitude. But at sufficiently large values this intrinsic noise can also affect the deterministic dynamics of the system.  Of particular interest is the phenomenon of stochastic resonance~\cite{benzi,zhou,gluc} (SR) in which  the addition of  random noise  amplifies the  pre existing subthreshold deterministic signal and  has been observed in many systems  such as physical~\cite{moss}, chemical~\cite{guderian}, electronics~\cite{postnov}, biological~\cite{moss2}and numerical model like FitzHugh–Nagumo (FHN)~\cite{gao}. Coherence resonance~\cite{pikovsky,kiss,santos} (CR), which is also called autonomous SR or internal signal SR, is an emergence of  regularity in the dynamics under the influence of purely stochastic perturbations. In CR, the maximum regularity is achieved at an optimum noise amplitude which has been observed in many experimental systems: optical~\cite{balle}, electrochemical~\cite{kiss}, chemical reactions~\cite{kenji},  electronic monovibrator circuit~\cite{postnov} as well as in numerical simulation of the standard model like FitzHugh–Nagumo (FHN) model~\cite{santos,pikovsky}, thermochemical system model~\cite{gorecki}.  The most important characteristic of CR is that the time scale of induced oscillations is determined  by the intrinsic dynamics of the system.

Plasma being a nonlinear and a complex medium with  numerous free energy sources which are expended by giving rise to several instabilities~\cite{merlino,klinger,pop:jaman}  which in turn interact to produce a background plasma noise  of broadband nature ranging from low frequency ion acoustic modes to high frequency electron plasma oscillations. The common sources of intrinsic noise in experimental systems also include the plasma fluctuations~\cite{fimo}, photons and fast neutrals in the system~\cite{scott}. Though they are small, they are widely used in estimating the plasma temperature and other parameters. The theme of this paper is to explore the effects of this kind of noise in generating coherent spikes when plasma is treated to be an excitable medium. There are several research works on the  external noise induced dynamics in the plasma system~\cite{jamman} as well as reports on the observation of CR in glow discharge plasma under the influence of external noise perturbation~\cite{jamman} but we have observed the same phenomena without any application of external noise.



In this paper, we present experimental investigations on intrinsic noise induced dynamics in a glow discharge plasma. Power spectrum  and normalized variance (NV)  have been used to analyze the time series of floating potential fluctuation. We have also estimated that the intrinsic noise level increases with the discharge voltage (DV) and hence  conjecture that it could be playing a  vital role in the observation of the CR phenomena as seen by the dip in the NV plot. The resonance phenomena has been also verified using the Hurst exponent and power spectrum.   To understand the experimental observations,  a FHN like model derived from an anharmonic oscillator equation for ion acoustic oscillations is considered. Since our experimental results reveal that the intrinsic noise  plays a role in the dynamics of the system, we included in the numerical model a forcing term consisting of a bias and a Gaussian noise.  The Gaussian noise is understood to play the role of the intrinsic noise present in the experimental system.

The rest of the paper is structured as follows: In Section II, we describe the experimental setup, followed by the results of the analysis of the floating potential fluctuations using normalized variance, Hurst exponent and power spectrum plot in Section III. In Section IV, we looked for a similar effect, i.e. CR phenomena, in a FHN like model. Conclusions and summary of the results  are presented in Section V.


\section{Experimental setup}
\label{section:Experimental setup}

\begin{figure}
  \centering
  \includegraphics[width=10cm]{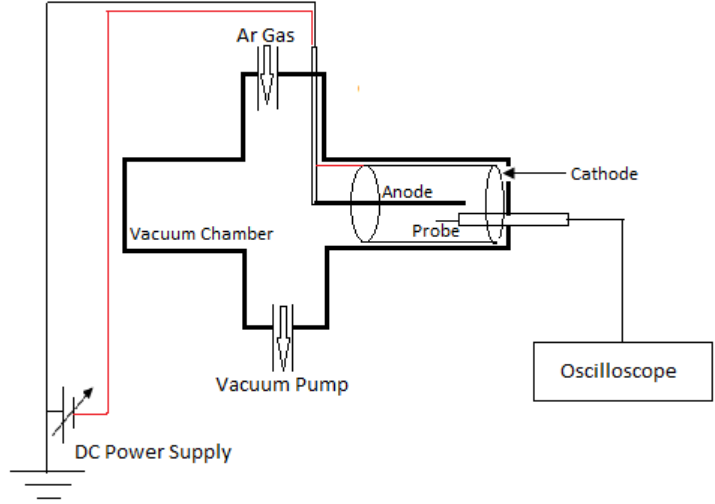}\\
  \caption{Schematic diagram of experimental setup}\label{setup}
\end{figure}

The schematic diagram of the  hollow cathode dc glow discharge plasma device~\cite{jamman} is shown in Fig. \ref{setup}. It has a cylindrical cathode of length and diameter $\sim 17$ $cm$ and $\sim 10$ $cm$ respectively and a central anode rod of diameter $\sim 1.6$ $mm$. The  operating pressure was about $.37$ $mbar$ and  the discharge voltage  could be varied from $0-1000$ $volts$.  The floating potential fluctuations were measured using a Langmuir probe connected to an oscilloscope.
The typical value of  electron temperature, ion temperature and density are estimated to be approximately $1-2$ $eV$, $0.1$ $eV$ and $10^9$  $cm^{-3}$ respectively.

\section{Result and Discussion}
\label{section:result}

The glow discharge plasma can exhibit a wide variety  of nonlinear phenomena depending upon the choice of operating pressure and discharge voltage (DV). For the purpose of the present experiments, we started the system at a discharge voltage $\sim$ $478 V$ where it exhibited fixed point behavior, and then the DV was increased monotonically and floating potential fluctuations were recorded.

\begin{figure}
  \centering
  \includegraphics[width=16cm]{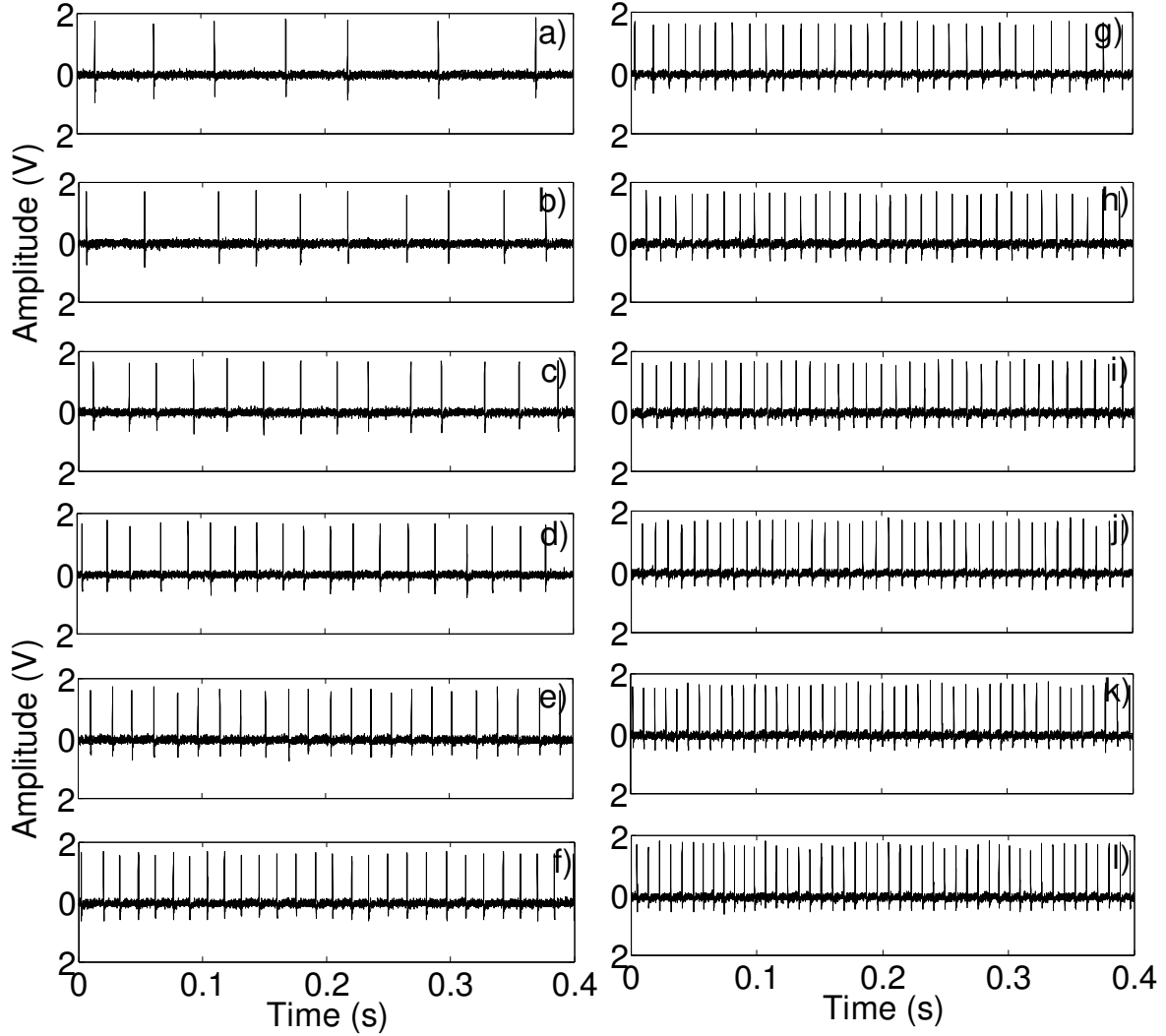}\\
  \caption{Time series of the floating potential fluctuation for different values of DV a) 479 V, b) 480 V, c) 481 V, d) 482 V, e) 483 V, f) 484 V, g) 485 V, h) 486 V, i) 487 V, j) 488 V, k) 489 V and l) 490 V. }\label{expraw}
\end{figure}

Fig. \ref{expraw}(a-j) shows the time series of the floating potential fluctuation at different values of DV as mentioned in the caption of the figure. The dynamics of the time series show that the system behaves as an excitable one, presenting characteristic spiking for values of DV when it exceeds a threshold. It is also seen from the Fig. \ref{expraw}(a-j) that the number of the spikes continuously increases with the increase in DV. The approximate values of the rise and the fall time of the spikes are 0.12 msec and 0.22 msec respectively. These time scales are comparable with the ion transit time scale ( $\tau = d/\sqrt{({{k_BT_i}\over {m}})}$ $\sim$ 0.1 msec, where $d$, $k_B$, m and $T_i$ are the electrode distance, Boltzmann constant, ion mass and ion temperature respectively. ) between two electrodes. So probably these spikes correspond to bunches of ions excited from the anode.

\begin{figure}
  \centering
  \includegraphics[width=10cm]{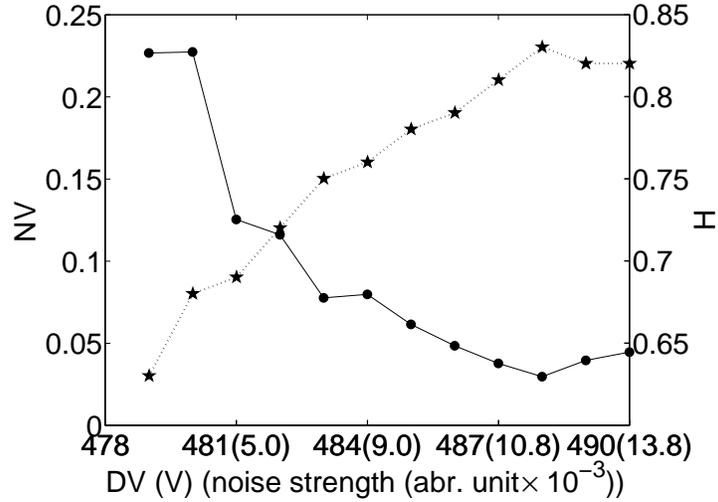}\\
  \caption{Plot of normalised variance (NV) (solid line) and the Hurst exponent (H) (dashed line) vs DV with its corresponding noise level given in the parenthesis.}\label{enor}
\end{figure}

\begin{figure}
  \centering
  \includegraphics[width=10cm]{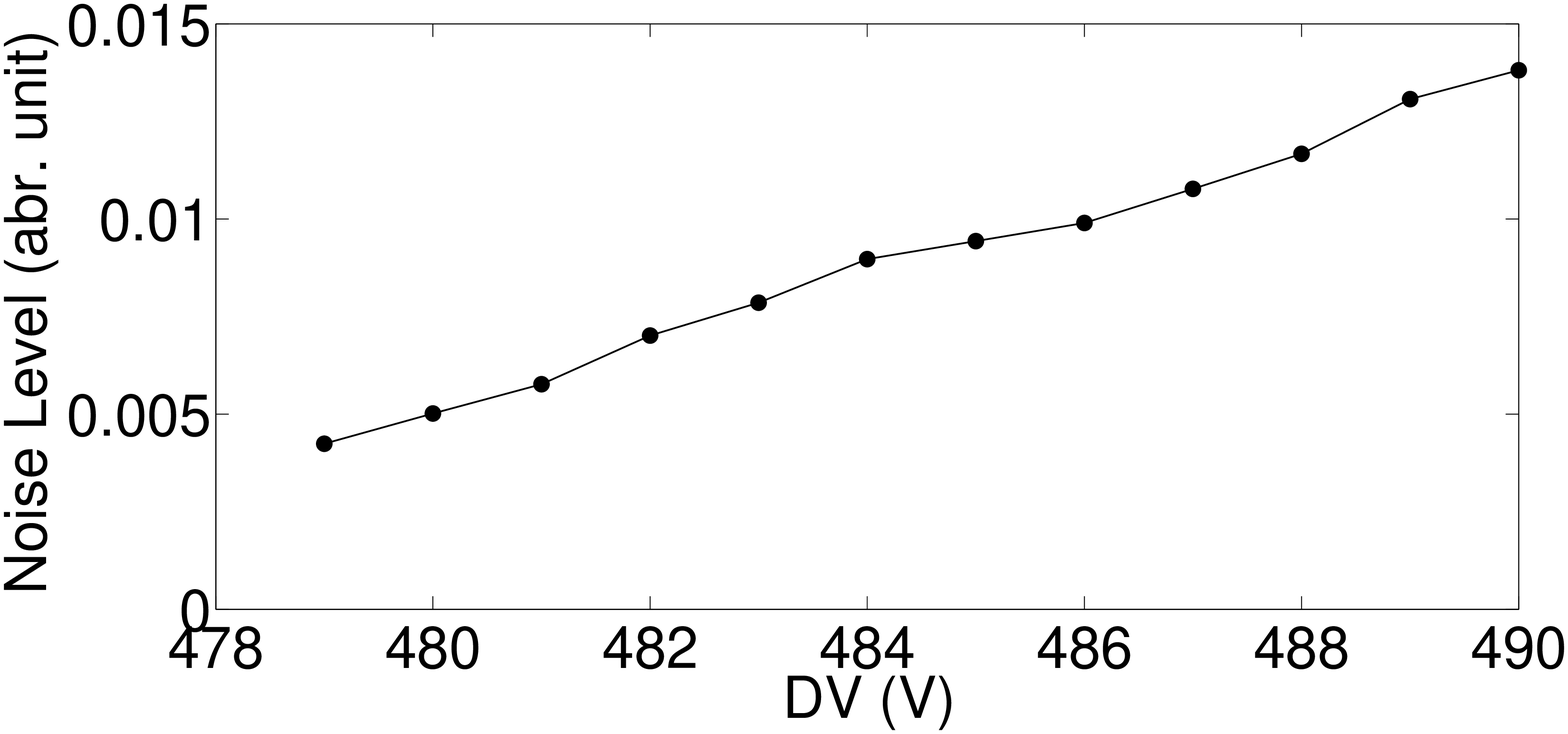}\\
  \caption{Intrinsic noise level as a function of DV}\label{noise}
\end{figure}

The normalized variance (NV) was used to quantify the  regularity in the spikes. It is defined as NV=std(ISI)/mean(ISI), where ISI is the time elapsed between successive spikes. It is evident that the value of the computed NV will be lower for the more regular induced dynamics. For purely periodic dynamics, the NV will be zero. Fig. \ref{enor} is the experimental NV curve as a function of DV. Higher value of NV at DV 481 V indicates the irregular nature of the  spikes. It is seen that NV decreases with the increase in the DV indicating the enhancement in regularity and maximum regularity is achieved at DV $\sim$ 488 V. Further increase in DV leads to irregularity in the system.  The minima in NV curve suggests that the phenomena is similar to coherence resonance. Generally, noise is responsible for the CR so we have estimated the intrinsic noise level with the help of wiener filter subroutine in matlab. Fig. \ref{noise} shows the intrinsic noise level as a function of DV depicting the enhancement in the intrinsic noise level with the increase in DV. As the intrinsic noise level increases with DV, it is clear that the regularity of the excitable spikes in the floating potential fluctuation increases with the increase in the noise level, upto an optimal value of the noise strength.

The estimation of characteristic correlation time $(\tau)$ using normalized autocorrelation function has been shown as one of the measures of coherent behavior~\cite{pikovsky,kenji}. The occurrence of a maxima in the $\tau$ vs noise amplitude curve at the point of maximum regularity has been reported  for the coherence resonance phenomena~\cite{pikovsky}. It is well known that the Hurst exponent can also be used as a measure of temporal correlation. Hence, to further characterize the coherence resonance behaviour, we evaluated  the Hurst exponent estimated  using rescaled range analysis (R/S) method~\cite{mandelbort}. The value of $H = 0.5, < 0.5, > 0.5$ and $1$ indicates the random, anti-correlated, correlated and periodic nature of the time series signal respectively. The dependence of the Hurst exponent on the DV and noise strength is shown in Fig. \ref{enor}. The nature of the H vs DV curve has an opposite trend with respect to NV vs DV curve. The plot clearly shows the coherence resonance maximum at DV $\sim$ 488 indicating the maximum temporal correlation at a particular discharge voltage and hence at a particular value of noise strength.

%

\begin{figure}
  \centering
  \includegraphics[width=16cm]{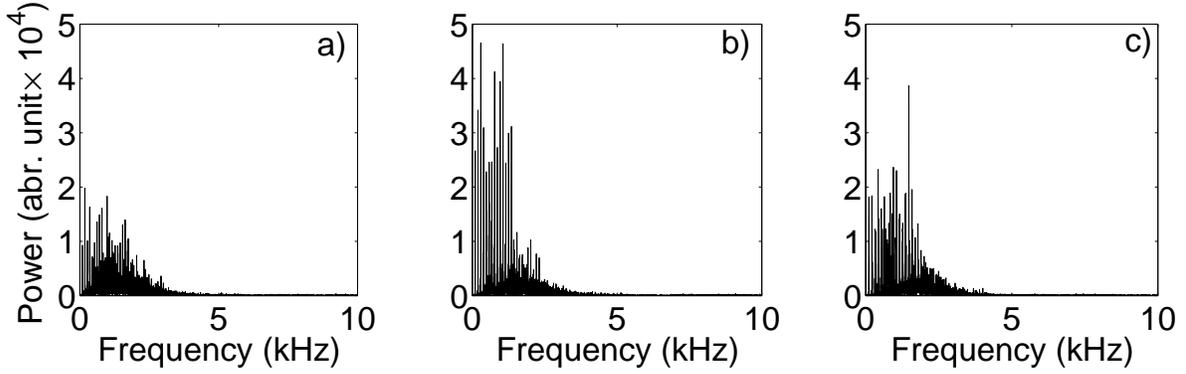}\\
  \caption{Power spectra for the floating potential fluctuation at DV a) 487 V, b) 488 V and c) 489 V.}\label{spectra}
\end{figure}

Figs. \ref{spectra}a, \ref{spectra}b and \ref{spectra}c show the power spectra of the floating potential fluctuation for DV 487 V, 488 V and 489 V respectively.   The power at DV  488 which is the point of coherence resonance shows a higher power than  the other values of DV. The average power of the  frequency band $(0-1500 Hz)$ in case Fig. \ref{spectra}b is approximately greater by a factor of $2.5$ and $1.7$ in comparison with Fig. \ref{spectra}a and Fig. \ref{spectra}c respectively.

\section{Numerical Simulation}
\label{section:simulation}
Keen et al.~\cite{keen} had derived an anharmonic oscillator equation for ion acoustic instabilities treating plasma as a two fluid model with source terms to contribute to the nonlinear effects. In an effort to understand the intrinsic noise induced CR in our experiments, we obtained  an excitable FHN like model from the  anharmonic oscillator equation  for ion acoustic instabilities~\cite{keen1,keen}. The anharmonic oscillator  for the ion density perturbation is given by
\begin{equation}\label{keen}
  \frac {d^{2}n_1}{dt^{2}}-(\alpha -2 \lambda n_1  - 3 \mu n_1^{2})\frac{dn_1}{dt}+ \omega_{0}^{2}n_{1}=0
\end{equation}
where $n_1$, $\alpha$, $\lambda$, $\mu$ and $\omega_0$ are the perturbed plasma density, ionization term, coefficient of two body recombination, coefficient of three body recombination and ion acoustic frequency respectively.

By normalizing the above equation using $ \tau=\omega t$; $x={n_1}/{n_0}$; $p= {\alpha/\omega}$; $q={2\lambda n_0/ \omega}$; $r={3\mu n_0^2/ \omega}$; $s={\omega_0^2/\omega^2}$  we obtain

\begin{eqnarray}\label{final}
 \ddot{x}-(p-qx-rx^2 )\dot{x}+sx= 0
 \label{sem}
\end{eqnarray}

By using a Li\'{e}nard-like coordinate, Eq. \ref{final} can be decomposed into the following system of two first order equation:

\begin{eqnarray}
  \dot{x} &=& (px-{{qx^2}\over {2}}-{{rx^3}\over {3}}) -sy\\
  \dot{y} &=& x
\end{eqnarray}

by rearranging the parameter, we have obtained a FHN like model given by
\begin{eqnarray}
\label{f1}
  \epsilon\dot{x} &=& (ax-{{bx^2}\over {2}}-{{cx^3}\over {3}}) -y\\
  \dot{y} &=& x \label{f2}
\end{eqnarray}
where $\epsilon, a, b$ and $c$ are $1/s, p/s, q/s$ and $r/s$ respectively.

The difference between the original FHN model and our present equation is the presence of the quadratic term.
Although there is no explicit externally applied noise present in the  experimental system, we consider the situation where noise is generated intrinsically. In order to investigate the behaviour of the nonlinear oscillations in the presence of an external discharge voltage, we include an additional biasing term $k$ and a noise term $r*\xi$ on the right hand side of Eq. \ref{f2}. $\xi$ is a Gaussian noise term and $r$ represent the strength of the noise.
\begin{eqnarray}
\label{f11}
  \epsilon\dot{x} &=& (ax-{{bx^2}\over {2}}-{{cx^3}\over {3}}) -y\\
  \dot{y} &=& x +k +r*\xi \label{f22}
\end{eqnarray}

This model does not  exactly represent the experimental system at a microscopic level, but we expect  to  describe the excitable dynamics of glow discharge plasma as this model looks similar to FHN model. Since the above phenomena were observed in  the small window of discharge voltage, we fixed the bias ($k$) at constant value such that the dynamics exhibit the excitable fixed point behavior  and varied the noise strength. The above Eqs. \ref{f11} and \ref{f22} are solved numerically using fourth order Runge Kutta method with initial conditions and time step $x=0$, $\dot y =1$ at $\tau =0$ and 0.01 respectively. The parameters $\epsilon$, $a,$ $b$, $c$ and $k$ are fixed at 0.01, 1, 0.95, 0.85 and 1.8 respectively.

\begin{figure}
  \centering
  \includegraphics[width=16cm]{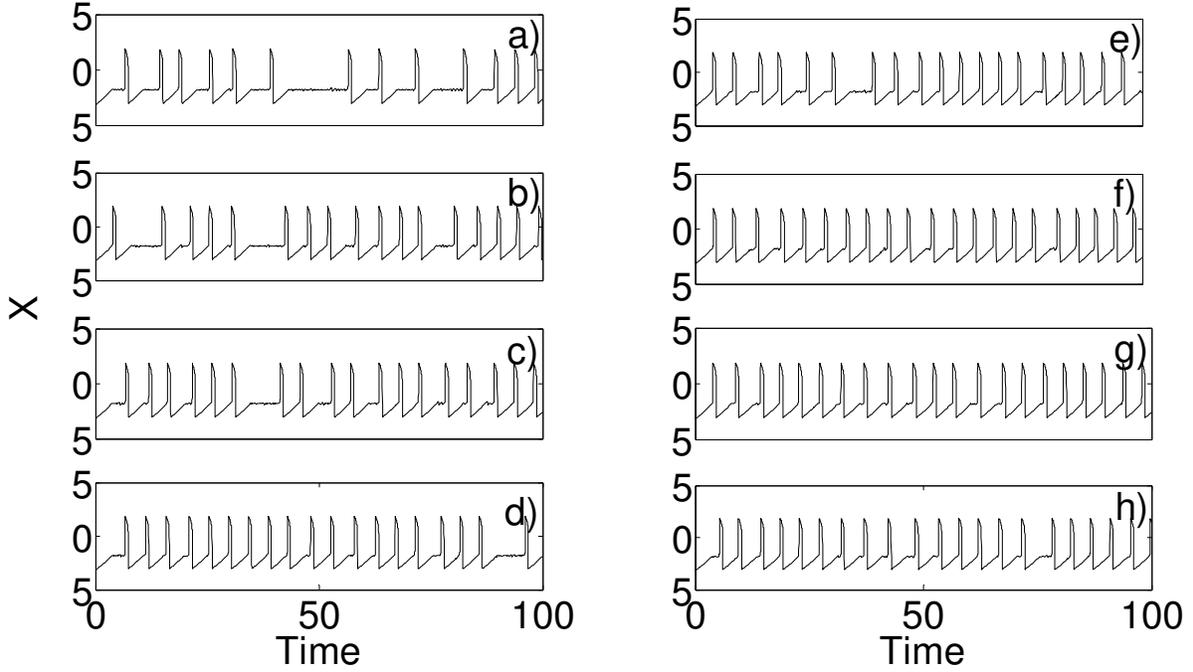}\\
  \caption{Solutions (x) of numerical model for various noise strength (r): (a) 0.10, (b) 0.15, (c) 0.20, (d) 0.25, (e) 0.30, (f) 0.35, (g) 0.40, (h) 0.45 and (i) 0.50. }\label{traw}
\end{figure}

Fig. \ref{traw} shows the spiking oscillation at various values of the noise strengths (r) as mentioned in the caption of the figure. It is observed that the number of spikes  increases with noise strength.

\begin{figure}
  \centering
  \includegraphics[width=10cm]{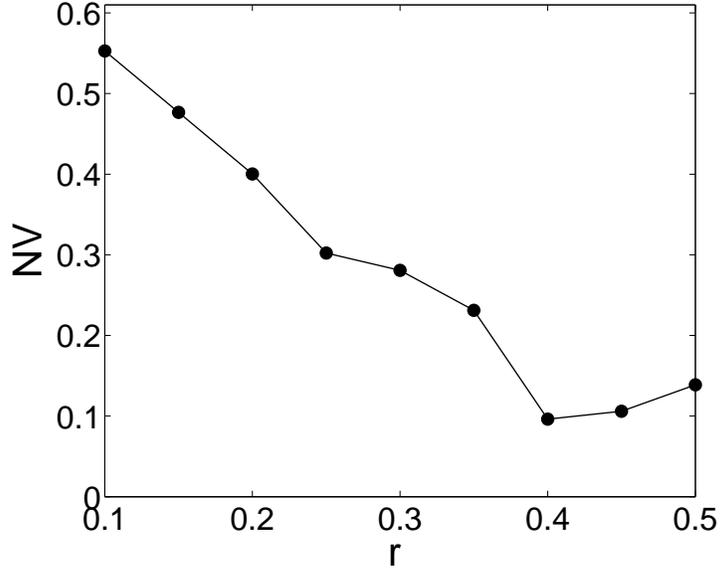}\\
  \caption{NV as a function of r}\label{tnor}
\end{figure}

Fig. \ref{tnor} shows the numerically computed NV curve.  The value of computed NV is higher at lower noise strength indicating the irregularity in spikes and  shows a minima at $r=0.4$ corresponding to an optimum noise level where the  maximum regularity of the generated spike sequence is observed. The numerically computed NV curve is consistent with the experimental results.

\begin{figure}
  \centering
  \includegraphics[width=16cm]{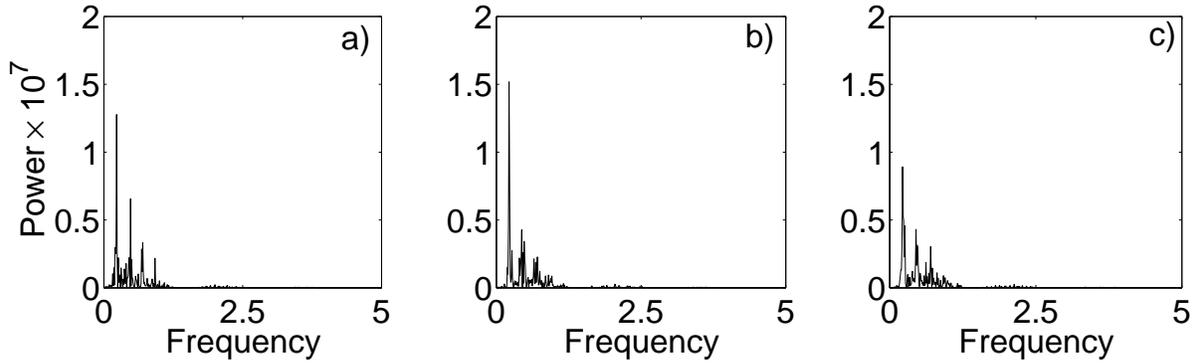}\\
  \caption{Power spectra of the numerical solution for noise strength (r) a) 0.35,  b) 0.4 and c) 0.5 V.}\label{tspectra}
\end{figure}

Power spectrum plot for  noise strength (r) 0.35, 0.4 and 0.5   are shown in Fig. \ref{tspectra}a, \ref{tspectra}b and \ref{tspectra}c respectively. It is seen that  the power is larger for the Fig. \ref{tspectra}b. Maximum power is observed for a particular noise level, which indicates the presence of  coherence resonance phenomena. This result also shows a good agreement with the result obtained from the numerically computed NV curve.

\section{Conclusion}
\label{section:conclusion}
Intrinsic noise induced coherence resonance has been observed in a glow discharge plasma. The amount of internal noise is dependent on the DV where an enhancement in noise level with discharge voltage has been observed.  The resonance curve (NV curve) is used to quantify the regularity of the intrinsic noise induced oscillation and hence to verify the CR phenomena.  It is shown that the floating potential fluctuations show maximum periodicity and maximum power in the power spectrum  for a particular DV i.e. at a particular value of intrinsic  noise. The utility of the Hurst exponent in the characterization of coherence resonance phenomena has been explored which suggests that Hurst exponent can be used a tool to identify the CR. In Refs. \cite{jamman} external noise induced CR  have been reported for glow discharge. In contrast, in the present work we presented that intrinsic noise induced CR is also possible in a glow discharge plasma.  A numerical model, resembling an FHN model obtained from anharmonic oscillator for ion acoustic instabilities, has been used to understand the dynamics of the observed  experimental results. The results obtained from the numerical simulation are in good agreement with that of experiments. It is quite likely  that noise induced spikes are related to the  bunches of ions emanating from the anode  and drift towards the cathode-anode gap. We hope our finding is helpful for studying the interaction of the intrinsic noise and plasma modes in plasma system and these results may be utilized to characterize various plasma based devices like plasma coating devices to improve their efficiency, plasma lasers to optimize the lasing output  where intrinsic noise can play a beneficial role.  Understanding the role of intrinsic noise therefore can be an important contribution to the study of excitable systems.

\section*{Acknowledgement}
The authors would like to acknowledge the director, SINP, for his constant support and Dipankar Das and Ashok Ram of the plasma physics division for the technical help.
\bibliographystyle{aipnum4-1}

\end{document}